\documentclass[lettersize,journal]{IEEEtran}
\usepackage{amsmath,amsfonts}
\usepackage{algorithmic}
\usepackage{algorithm}
\usepackage{array}
\usepackage[caption=false,font=normalsize,labelfont=sf,textfont=sf]{subfig}
\usepackage{textcomp}
\usepackage{stfloats}
\usepackage{url}
\usepackage{verbatim}
\usepackage{graphicx}
\usepackage{cite}
\usepackage{amssymb}
\usepackage{color}

\usepackage{upgreek}

\usepackage{flushend,cuted}

\allowdisplaybreaks[4]

\begin{document}

\title{Dynamic Multi-time Scale User Admission and Resource Allocation for Semantic Extraction in MEC Systems}

\author{Yuanpeng Zheng,~\IEEEmembership{Student Member,~IEEE,} Tiankui Zhang,~\IEEEmembership{Senior Member,~IEEE,} Jonathan Loo
\thanks{}
\thanks{}
\thanks{
This work was supported by National Natural Science Foundation of China under Grants 61971060.
(Corresponding author: Tiankui Zhang)}
\thanks{Yuanpeng Zheng, Tiankui Zhang are with the
School of Information and Communication Engineering,
Beijing University of Posts and Telecommunications, Beijing 100876, China (e-mail: \{zhengyuanpeng, zhangtiankui\}@bupt.edu.cn).}
\thanks{Jonathan Loo is with the School of Computing and Engineering, University of West London,  London W5 5RF, U.K. (e-mail: jonathan.loo@uwl.ac.uk).}
}

\markboth{}%
{Shell \MakeLowercase{\textit{et al.}}: A Sample Article Using IEEEtran.cls for IEEE Journals}

\IEEEpubid{ }

\maketitle

\begin{abstract}
  This paper investigates the semantic extraction task-oriented dynamic multi-time scale user admission and resource allocation in mobile edge computing (MEC) systems. Amid prevalence artificial intelligence applications in various industries, the offloading of semantic extraction tasks which are mainly composed of convolutional neural networks of computer vision is a great challenge for communication bandwidth and computing capacity allocation in MEC systems. Considering the stochastic nature of the semantic extraction tasks, we formulate a stochastic optimization problem by modeling it as the dynamic arrival of tasks in the temporal domain. We jointly optimize the system revenue and cost which are represented as user admission in the long term and resource allocation in the short term respectively. To handle the proposed stochastic optimization problem, we decompose it into short-time-scale subproblems and a long-time-scale subproblem by using the Lyapunov optimization technique. After that, the short-time-scale optimization variables of resource allocation, including user association, bandwidth allocation, and computing capacity allocation are obtained in closed form. The user admission optimization on long-time scales is solved by a heuristic iteration method. Then, the multi-time scale user admission and resource allocation algorithm is proposed for dynamic semantic extraction task computing in MEC systems. Simulation results demonstrate that, compared with the benchmarks, the proposed algorithm improves the performance of user admission and resource allocation efﬁciently and achieves a ﬂexible trade-off between system revenue and cost at multi-time scales and considering semantic extraction tasks.
\end{abstract}


\section{Introduction}
In recent years, mobile edge computing (MEC), which supports not only computing but also communications and storage, has become a key technology to solve many related problems with specific requirements\cite{ref1,ref2}. By being closer to the edge of network than traditional cloud computing systems, MEC can obviously improve the quality of user experience, including optimization of delay and energy consumption\cite{ref3}. Devices can signiﬁcantly reduce their response times and energy consumption by ofﬂoading the computing tasks to nearby edge network, hence resource capacity and scheduling of MEC systems become a very important issue, especially in the context of the increasing number of intelligent tasks. The rapid development of network edge applications such as the Internet of Things (IoT) indicates the change of service requirements and diversity of tasks, nevertheless, few existing works consider the various performance requirements of these dynamics applications\cite{ref4} and the characteristics of computing tasks\cite{ref5}. Obviously, how to efficiently allocate resource to support the dynamics demand of services is still an unaddressed problem.

Hence, in the context of massive IoT devices deployment, limited terminal computing capacity and battery capacity, and increasingly complex computing tasks, existing works on resource allocation in MEC systems have become specific and multidimensional\cite{ref6,ref7,ref8,ref9,ref10,ref30,ref31}. S. Zarandi \textit{et al.}\cite{ref6} investigated a way of combining MEC and network slicing, and proposed the optimization of the weighted sum of the difference between the observed delay and the delay requirement. By considering edge users and large data volume, a power consumption and delay optimization problem in unmanned aerial vehicle (UAV) assisted MEC systems was addressed by G. Faraci \textit{et al.}\cite{ref7}. As a key scenario, an efficient method of MEC and network slice integration which was deployed on the IoT platform was proposed by J. Y. Hwang \textit{et al.}\cite{ref8} to maximize the effect of decreasing delay and traffic prioritization. X. Cao \textit{et al.}\cite{ref9} introduced a new MEC setup where a UAV was served by cellular ground base stations for computation ofﬂoading to minimize the UAV's task completion time considering computing capacity. Considering computing tasks, the MEC technique combined with network slicing and non-orthogonal multiple access was leveraged by M. A. Hossain \textit{et al.}\cite{ref10} to minimize the total latency of the computing tasks with energy constraints. T. Zhang \textit{et al.} \cite{ref30} considered that a UAV equipped with an MEC server was deployed to serve a number of terminal devices of Internet of Things in a finite period, which aimed to minimize the total energy consumption including communication-related energy, computation-related energy and UAV's flight energy by optimizing the bits allocation. J. Feng \textit{et al.} \cite{ref31} proposed a heterogeneous computation and resource allocation framework based on a heterogeneous mobile architecture to achieve effective implementation of federated learning. Obviously, the above works do not consider the influence of the stochastic nature and specific computing capacity consumption of computing tasks on user admission and resource allocation. The research on specific tasks have become important in MEC systems considering communication bandwidth and computing capacity allocation. Ignoring the characteristics will result in some errors in real scenario and hence can not satisfy delay requirements well.

The rise of semantic communication research in recent years has brought more requirements to the MEC field. Meanwhile, it proposes more scenarios of specific tasks in MEC systems and a few studies on quantification of computational complexity of intelligent tasks has been discussed\cite{ref11,ref12,ref13,ref14,ref15,ref16}. Some works proposed practical schemes to deploy semantic communication to MEC systems. H. Xie \textit{et al.}\cite{ref11} proposed a brand new framework of semantic communication where a deep learning based semantic communication system for text transmission combined with deep learning, natural language processing and semantic layer communication was constructed. After that, H. Xie \textit{et al.}\cite{ref12} considered a semantic communication system which was constructed between edge and IoT devices where MEC servers trained and updated the semantic communication model based on deep learning, and the IoT devices collected and transmitted data based on the training model. H. Qi \textit{et al.}\cite{ref13} investigated a model named PALEO which was applied to analyze performance of deep neural network (DNN). Nevertheless, D. Justus \textit{et al.}\cite{ref14} indicated that the presentation of computational complexity of PALEO was not accurate because of many other influence factors, and proposed an alternative strategy which predicted execution time by training a deep learning network including network features and hardware features. An approximation strategy of optimization of DNN training was proposed by D. Bienstock \textit{et al.}\cite{ref15}, which modelled DNN as a directed graph to control approximation error of computational complexity. M. Bianchini \textit{et al.}\cite{ref16} proposed a new approach to study how the depth of feedforward neural networks impacted on their ability to implement high complexity functions and indicated how the complexity depended on the number of hidden units and the used activation function. It is shown that the resource allocation problems for the semantic communication including the communication bandwidth and the computing capacity with the dynamic arrival of task in MEC systems has not been fully studied and it is hard but important to present characteristics and complexity of intelligent computing tasks.

There are still some studies considering the dynamic of mobile network and the stochastic nature of computing tasks\cite{ref5,ref17,ref18,ref19}. F. Guo \textit{et al.}\cite{ref17} designed the framework where the service requirements of some IoT applications were changing. In\cite{ref18}, Y. Xiao \textit{et al.} considered the dynamic of fog computing networks to maximize the utilization efﬁciency of available resources while balancing the workloads among fog nodes. The real-time dynamics of the network resource requests have been discussed in \cite{ref19} by N. Van Huynh\textit{et al.} and obtained the optimal resource allocation policy under the dynamics of the frequency of request. J. Feng \textit{et al.}\cite{ref5} considered the stochastic nature of tasks and proposed an architecture that maximized revenue of network providers in MEC systems, where a multi-time scale scheme was adopted to increase revenue on the basis of QoS guarantee. However, it is necessary to integrate the stochastic nature of tasks and quantification of complexity of computing tasks in MEC systems. Semantic extraction tasks which are mainly composed of convolutional neural networks (CNN) of computer vision gradually become the mainstream on dynamic resource allocation. In conclusion, modelling dynamic and computing capacity of semantic extraction tasks in MEC systems has not been considered yet according to the above works.

\subsection{Motivation and Contribution}
As mentioned above, the combination of dynamic multi-time admission and resource allocation in MEC systems with specific computing tasks, i.e., semantic extraction tasks, is still an unaddressed research area, which motivates this contribution. In this paper, we formulate a stochastic optimization problem by modelling it as dynamic arrival of tasks in temporal domain considering the stochastic nature of the semantic extraction tasks. In order to investigate the dynamic arrival and stochastic nature of tasks, we adopt multi-time scale to represent traffic variations. Based on the dynamic model, we optimize the average utility over time that consists of the system revenue and cost which depends on user admission in the long term and resource allocation in the short term. We also model computing characteristic of semantic extraction task as a formula based on the structure of CNN. The primary contributions of this paper are as follows:

\begin{itemize}
\item{We formulate a stochastic optimization problem for dynamic user admission and resource allocation considering the stochastic nature of semantic extraction tasks in MEC systems. We set up a queue model to represent dynamic of semantic extraction tasks and define the operator's utility which consists of long-time-scale revenue depending on the number of users and short-time-scale cost depending on power consumption in order to achieve continuous revenue in temporal domain with as little cost as possible at each time slot. For this study, we adopt a formula based on the structure of CNN to quantify the relationship between input data and computational complexity of semantic extraction tasks.}
\item{We solve the highly coupled problem without any prior knowledge of traffic distributions or channel information with the assistance of the Lyapunov optimization with maximization of the number of users and minimization of power consumption.
We decouple manifold optimization variables on the dimension of time scale and propose a multi-time scale user admission and resource allocation algorithm for semantic extraction tasks where the dynamic user admission subproblem is in the long term and user association subproblem, bandwidth allocation subproblem and computing capacity allocation subproblem are in the short term. The dynamic user admission subproblem in the long term is settled by a heuristic iteration method and resource allocation subproblems in the short term are solved in closed forms.}
\item{We demonstrate the simulation results which verify that our framework is applicable to semantic scenario in MEC systems and the proposed algorithm has significant effect for the multi-time scale problem solving. It is shown that, compared with the benchmarks, the proposed algorithm improves the performance of user admission and resource allocation efficiently and achieves a flexible trade-off between system revenue and cost at multi-time scales and considering semantic extraction tasks.}
\end{itemize}

\subsection{Organization}
The rest of this paper is organized as follows. In Section II, we introduce system model and problem formulation. In Section III, we decompose the coupling problem into resource allocation in the short-time slot and user admission in the long-time slot. The performance of the proposed algorithm is evaluated by the simulation in Section IV, which is followed by our conclusions in Section V.

\section{Stochastic Optimization Problem Formulation}
We consider that fog radio access network (F-RAN) is built on MEC systems, and communication and computing between terminals and MEC are for specific semantic extraction tasks, as shown in Fig. 1. We equip MEC servers on small base stations (SBS) to form MEC systems, which is assembled as $K^{S}$ = \{1,\,...,\,\textit{k},\,...,\,\textit{K}\}. In order to consider dynamic allocation of resources in temporal domain to dynamically meet the demands of multiple task slices, we design two types of time slots based on the time-slotted system where one is a long time slot (LTS) and the other is a short time slot (STS). In this paper, our system contains multiple LTSs which are dedicated to user admission and the length of the LTS is \textit{T}. We assume that each LTS contains \textit{p} STSs which are dedicated to resource allocation and the length of STS is $\tau$, i.e., $T = p\tau$. 
At LTS \textit{l}, we denote the set of users by $U^{S}$ = \{1,\,...,\,\textit{u},\,...,\,\textit{U}\}, and the set of specific tasks by $M^{S}$ = \{1,\,...,\,\textit{m},\,...,\,\textit{M}\}. 
At the beginning of each LTS the network operator can decide user admission and at the beginning of each STS resource allocation policies is given. Let the admission control variable of the user \textit{u} accessing MEC systems be $y_u(l)\in\{0,1\}$, where $y_u(l) = 1$ denotes user \textit{u} is admitted by MEC systems and $y_u(l) = 0$ means the opposite. The multi-time scale system will be discussed in detail in the following sections of this section. Let the bandwidth resource of each SBS be $W_k$, computing capacity of each MEC be $F_k$. The delay limit for semantic extraction tasks is set to $\tilde{t}_m$. 

\begin{figure}[!t]
  \centering
  \includegraphics[scale=0.63]{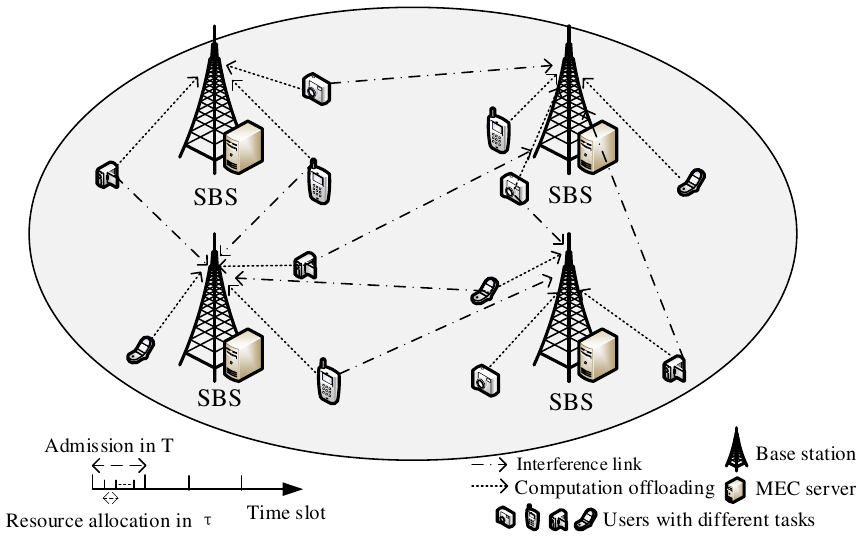}
  \caption{The system scenario.}
  \label{fig_1}
\end{figure}

\begin{table}[h]
  \renewcommand\arraystretch{1.5}
  \caption{Main Symbol and Variable List}
  \centering
  \begin{tabular}{|p{1.3cm}|p{6cm}|}
  \hline
  Notation & Description \\
  \hline
  $K,U,M$ & Number of SBSs, users and tasks\\
  \hline
  $T,\tau$ & The long and short time slot\\
  \hline
  $W_k$ & Bandwidth resource of each SBS\\
  \hline
  $F_k$ & Computing capacity of each MEC\\
  \hline
  $\tilde{t}_m$ & The delay limit for semantic extraction tasks\\
  \hline
  $B_{bus}$ & The bus bandwidth of the hardware devices within the SBS\\
  \hline
  $\mathcal{K}$ & The effective switched capacitance of the MEC server\\
  \hline
  $\eta$ & The parameter for adjusting the overhead weights\\
  \hline
  $y_u$ & Indicator of whether user $u$ admitted by MEC systems\\
  \hline
  $x_{uk}$ & Indicator of whether user $u$ accessing SBS $k$\\
  \hline
  $z_{um}$ & Indicator of whether task $m$ requested by user $u$\\
  \hline
  $r_{uk}$ & The uplink transmission rate of user $u$ accessing SBS $k$\\
  \hline
  $w_{uk}$ & The uplink bandwidth resource obtained by user $u$ accessing SBS $k$\\
  \hline
  $f_{uk}$ & The computing capacity allocated by SBS $k$ for user $u$\\
  \hline
  $t_u^{comm}$ & The transmission delay of the raw data collected by user $u$ in the wireless link\\
  \hline
  $t_u^{comp}$ & The computing latency incurred by user u to perform tasks on SBS\\
  \hline
  $t_u^{bus}$ & The latency generated by the bus transfer of data between the hardware within the system\\
  \hline
  $a_u$ & The raw data that user $u$ connected to SBS $k$ collects\\
  \hline
  $A_u$ & The random arrival process of tasks of user $u$\\
  \hline
  $\mathbf{\Upphi_I}$ & The amount of tasks in the current queue that need to be unloaded\\
  \hline
  $\mathbf{\Upphi_{II}}$ & The number of tasks currently in the bus transfer\\
  \hline
  $\mathbf{\Phi}$ & The number of tasks offloaded but unprocessed in cache\\
  \hline
  $F_{um}$ & The required computing resource of task $m$ of user $u$\\
  \hline 
  $P_{uk}$ & The computational power consumption of SBS $k$ to handle the user $u$ offloading task\\
  \hline 
  $v_{u}$ & The access control weighting parameter determined by the task delay limit\\
  \hline
  $G_L,G_S$ & The long-term revenue and short-term cost\\
  \hline
\end{tabular}
\end{table}

\subsection{Communication Model}
In our system, we adopt a more convenient communication model\cite{ref20} which can be easily modified to other general models to complete our design. At STS \textit{t}, the indicator variable for user \textit{u} accessing to SBS \textit{k} is denoted by $x_{uk}\in\{0,1\}$. Assume that a user can only access one SBS in a short time slot, then the uplink transmission rate of user \textit{u} accessing SBS \textit{k} is given by
\begin{equation}
  \label{deqn5}
  r_{uk}(t) = w_{uk}(t){\log}_2\left(1+\frac{p_ug_{uk}(t)}{I_{uk}(t)+\sigma^2 }\right),
\end{equation}
where $w_{uk}(t)$ is uplink bandwidth resource obtained by user \textit{u}, $g_{uk}(t)$ is channel gain between SBS \textit{k} and user \textit{u}, $p_u$ is the transmit power from user \textit{u} to SBS \textit{k}, and $I_{uk}(t)$ is the co-channel interference from users connected to other SBS, i.e. $I_{uk}=\sum_{i\in K^S,i\neq k} {g_{ui}(t)}p_u$. $\sigma^2$ is the noise power. We denote the raw data that user \textit{u} connected to SBS \textit{k} collects as $a_u(t)$, such as pictures of the industrial environment that need to be semantically segmented and given instructions. The raw data is transmitted to SBS through the uplink channel. Therefore the transmission delay of the raw data collected by user \textit{u} in the wireless link is
\begin{equation}
  \label{deqn6}
  t^{comm}_u(t)=\frac{a_u(t)}{r_u(t)},
\end{equation}
where $r_u(t)= \sum_{k=1}^{K} {x_{uk}(t)} r_{uk}(t)$.

{In our model, we let $A_u(t)$ denote the random arrival process of tasks of user \textit{u} in each STS \textit{t}. For processing convenience, we assume that $A_u(t)$ is independently and identically distributed between STSs and $\mathbb{E} \{A_u(t)\}=\lambda$ for all STSs.} Let ${\mathbf{\Upphi_I}}(t)$ denote the amount of tasks in the current queue that need to be unloaded. The dynamics of the task offloading queue is given by
\begin{equation}
  \label{deqn7}
  \begin{split}
    \mathbf{\Upphi_I}(t+1)=& \max \left\{{\mathbf{\Upphi_I}}(t)-\tau \boldsymbol{y}(l)\cdot \boldsymbol{r}(t) ,0\right\} + \boldsymbol{y}(l)\cdot \boldsymbol{A}(t),
  \end{split}
\end{equation}
where $\max$ represents queue accumulation of ${\mathbf{\Upphi_I}}$ exists only when the queue arrival is greater than the queue departure, otherwise it is 0, and $\boldsymbol{y}(l)$,  $\boldsymbol{r}(t)$ and $\boldsymbol{A}(t)$ are the vector representations of $y_u(l)$, $r_u(t)$ and $A_u(t)$.

At STS \textit{t}, user \textit{u} transmits its collected raw data $a_u(t)$ such as various captured images, etc., to the SBS for a specific semantic extraction task \textit{m} to generate semantic extracted feature data. Then the MEC feeds the semantic data back to the user for further operation via the downlink channel. In the context of the image semantic segmentation task in our scenario, the feature data is extremely small compared to raw data $a_u(t)$, therefore the downlink transmission delay can be neglected.

\subsection{Semantic Extraction Task}
The design of semantic extraction task oriented MEC systems becomes increasingly important as intelligent tasks become mainstream, especially lightweight semantic communication network combined with IoT\cite{ref12}. In this paper, we consider some image recognition applications of industrial Internet where the image semantic extraction algorithm based on CNN is mainly used. The computational complexity of those applications primarily depends on not only input raw data but also CNN. Semantic extraction tasks in scenario of MEC systems needs to be considered separately from the general task to this extent. 

In our system, we design a computing model for semantic extraction tasks which is specific to CNN. The required computing resource of CNN is determined by the amount of data and model parameters associated with the input of the convolutional layer, and the network model parameters are task-specific\cite{ref13}. We denote the model parameter of task \textit{m} as $n_m$, and the specific value is determined by the number of filters of CNN. Therefore, for ease of representation, the basic computing model for semantic extraction tasks at STS \textit{t} is expressed as
\begin{equation}
  \begin{split}
    \label{deqn8}
    F_{um}(a_u(t))=&n_m a_u(t) +\\
    &{\log}\left(\frac{a_u(t)}{3N}\right) \left(\frac{n_ma_u(t)}{N}+a_u(t)+\frac{n_ma_u(t)}{3}\right) 
  \end{split}
\end{equation}
The above equation represents the approximate function of raw data volume (Byte) and the required computing resource (Gigacycle), where the constant 3 is the number of channels and \textit{N} is the number of input feature maps. 
Therefore, we set the required computing resource of user $u$ as
\begin{equation}
  \label{deqn8_1}
  F_{u}(a_u(t)) = \sum_{m=1}^M {z_{um}(t)}F_{um}(a_u(t)),
\end{equation}
which means the computing resource needed to process the task of user $u$.

\subsection{Computing Model}
We propose a complete system-level computing model for semantic extraction tasks. Let the indicator variable for task \textit{m} of user \textit{u} at STS \textit{t} be $z_{um}(t)\in\{0,1\}$, where $z_{um}(t) = 1$ means the task of user \textit{u} is \textit{m} and $z_{um}(t)=0$ is the opposite. $z_{um}(t)$ is known as the indicator of user's request and $\sum_{m=1}^M z_{um}(t)=1$. We set the computing capacity that is allocated by SBS \textit{k} for user \textit{u} as $f_{uk}(t)$. Computing latency incurred by user \textit{u} to perform tasks on SBS is given by
\begin{equation}
  \label{deqn9}
  t^{comp}_u(t)=\frac{F_{u}(a_u(t))}{f_u(t)},
\end{equation}
where $f_u(t) = \sum_{k=1}^{K}x_{uk}(t)f_{uk}(t)$ (Gigacycle/s), which represents the computing capacity that MEC systems allocate to user $u$ to process offloading tasks.

At STS \textit{t}, the processing of computing tasks also requires consideration of the latency generated by the bus transfer of data between the hardware within the system\cite{ref13,ref14}, which is denoted as
\begin{equation}
  \label{deqn10}
  t^{bus}_u(t)=\frac{a_u(t)}{B_{bus}},
\end{equation}
where $B_{bus}$ represents the bus bandwidth of the hardware devices within the SBS. Therefore, the total delay for user \textit{u} to access SBS \textit{k} to complete the task processing is given by
\begin{equation}
  \label{deqn11}
  t_u(t)=t^{comm}_u(t)+t^{comp}_u(t)+t^{bus}_u(t).
\end{equation}

We consider the computational power consumption of SBS \textit{k} to handle the user \textit{u} offloading task, which is expressed as
\begin{equation}
  \label{deqn12}
  P_{uk}(t) = \mathcal{K}_{esc}f_{uk}^3(t), 
\end{equation}
where $\mathcal{K}_{esc}$ is the effective switched capacitance of the MEC server. Then the total power consumption of system is 
\begin{equation}
  \label{deqn13}
  P(t) = <\boldsymbol{x}(t), \boldsymbol{\mathcal{P}}(t)>,
\end{equation}
where $\boldsymbol{x}(t)$ and $\boldsymbol{\mathcal{P}}(t)$ are the matrix representations of $x_{uk}(t)$ and $P_{uk}(t)$ and $<\cdot,\cdot>$ represents matrix inner product.

In our model, we consider transmission between multiple hardware connected by bus inside the MEC server. This type of transmission can also have an impact on the task queue, therefore we model this impact as the bus transfer queue. The bus transfer queue is after the task offloading queue and is independent of it. Let ${\mathbf{\Upphi_{II}}}(t)$ denote the number of tasks currently in the bus transfer, the dynamics of bus transfer queue is expressed as
\begin{equation}
  \label{deqn14}
  \begin{split}
    \mathbf{\Upphi_{II}}(t+1)=&\max\left\{{\mathbf{\Upphi_{II}}}(t)-\sum_{u=1}^U {y_u(l)}B_{bus}\tau ,0\right\} + \\
    & \min\left\{ \boldsymbol{y}(l)\cdot \boldsymbol{r}(t),{\mathbf{\Upphi_I}}(t)\right\},
  \end{split}
\end{equation}
where $\max$ represents queue accumulation of ${\mathbf{\Upphi_{II}}}$ exists only when the queue arrival is greater than the queue departure, otherwise it is 0, $\min$ represents the effect of queue arrival and queue accumulation of ${\mathbf{\Upphi_I}}$ on accumulation of ${\mathbf{\Upphi_{II}}}$ in tandem queue and $\boldsymbol{f}(t)$ is the vector representation of $f_u(t)$.

Thereafter computing tasks are offloaded to the MEC for processing. Assuming that there is sufficient cache in MEC systems to store offloaded but unprocessed tasks, the dynamics of the computational processing queue is given by
{
\begin{equation}
  \label{deqn15}
  \begin{split}
    \mathbf{\Phi}(t+1)=&\max\left\{\mathbf{\Phi}(t)-\boldsymbol{y}(l)\cdot \boldsymbol{f}(t) ,0\right\}+\\
    &\min\Bigg\{\sum_{u} F_{u}(y_u(l)B_{bus})(t),\sum_{u}F_{u}({\mathbf{\Upphi_{II}}}(t))\Bigg\},
  \end{split}
\end{equation}
}
where $\max$ represents queue accumulation of $\mathbf{\Phi}$ exists only when the queue arrival is greater than the queue departure, otherwise it is 0, and $\min$ represents the effect of queue arrival and queue accumulation of ${\mathbf{\Upphi_{II}}}$ on accumulation of $\mathbf{\Phi}$ in tandem queue. 

\subsection{Utility Model and Problem Formulation}
In this paper, we consider the trade-off between the revenue and the cost of the optimization system, where the revenue depends on the number of admitted users associated with the admitted control weighting parameter and the cost depends on the computational energy consumption. The admitted control weighting parameter determined by the importance of users to revenue is expressed as $v_u$, which is given by

\begin{equation}
  \label{deqn16}
  v_u(l) = \frac{ {\sum\limits_{t=pl}^{p(l+1)-1}} \sum\limits_{m=1}^{M} z_{um}(t) \tilde{t}_m}{T},
\end{equation}
which is seen as the importance distribution for users at LTS \textit{l} determined by the average task delay limit. Then the revenue expression is
\begin{equation}
  \label{deqn17}
  G_L(l) = \boldsymbol{y}(l) \cdot \boldsymbol{v}(l),
\end{equation}
where $\boldsymbol{v}(l)$ is the vector representation of $v_u(l)$, $G_L(l)$ expresses that the influence of admitted users to revenue is determined by weighting parameter $v_u$. We investigate that computing energy cost of the system in long time slot $T$ is 
\begin{equation}
  \label{deqn17-1}
  G_S(l)= {\sum_{t=pl}^{p(l+1)-1}} {P(t)}.
\end{equation}

In that case the system utility is expressed as
\begin{equation}
  \label{deqn18}
  G(l) = G_L(l)- \eta G_S(l),
\end{equation}
\textbf{Remark 1.} \textit{From (\ref{deqn17})(\ref{deqn17-1}), we notice that $\eta$ is the parameter for adjusting the revenue and cost weights. Hence, different values of $\eta$ may effect the trade-off between the revenue in LTS and the cost in STS and stabilize the system utility. However, other comparison algorithms do not have this characteristic of balancing revenue and cost because of different treatment methods of admission and resource allocation. Therefore, our proposed algorithm can perform well under different values of $\eta$ in (\ref{deqn18}).}

Furthermore, we denote the average utility as
\begin{equation}
  \label{deqn19}
  \overline{G} = \lim_{Z \to \infty} {\frac{1}{Z}} \sum_{l=0}^{Z-1}{\mathbb{E}}\{G(l)\},
\end{equation}
where $\overline{G}$ represents the average utility of system on all time slots and is for constructing Lyapunov stochastic optimization problem in the following.

According to our model above, we investigate the operator's utility maximization problem in MEC systems by jointly controlling system admission ${\boldsymbol{y}}(l)$, user association ${\boldsymbol{x}}(t)$, bandwidth allocation ${\boldsymbol{w}}(t)$ and computing capacity allocation ${\boldsymbol{f}}(t)$. In particular, we formulate it as the following stochastic optimization problem.
\begin{equation}
  \label{deqn20}
  \begin{split}
    &\max_{{\boldsymbol{y}}(l),{\boldsymbol{x}}(t),{\boldsymbol{w}}(t),{\boldsymbol{f}}(t)}\overline{G}\\
    &s.t. \ \; (\text{C}1): y_u(l)\in \{0,1\},\forall u,l,\\
    &\qquad (\text{C}2): x_{uk}(t)\in \{0,1\},\forall u,k,t,\\
    &\qquad (\text{C}3):\sum_{k=1}^K {x_{uk}} \leqslant 1,\forall u,t,\\
    &\qquad (\text{C}4): y_{u}(l)t_u(t) \leqslant \sum_{m=1}^{M} z_{um}(t)\tilde{t}_m,\forall u,t,\\
    &\qquad (\text{C}5): \sum_{u=1}^U x_{uk}(t)w_{uk}(t) \leqslant W_k,\forall k,t,\\
    &\qquad (\text{C}6): \sum_{u=1}^U x_{uk}(t)f_{uk}(t) \leqslant F_k,\forall k,t,\\
    &\qquad (\text{C}7): \overline{Q}_I < \infty, \overline{Q}_{II}<\infty, \overline{{\mathbf{\Phi}}}<\infty,\forall t.
  \end{split}
\end{equation}

In \eqref{deqn20}, (C2) and (C3) indicate that a user can only access one SBS. (C4) is the delay requirements of tasks. (C5) and (C6) denote the limit of bandwidth and computing capacity of each SBS. (C7) represents that the data rate should be greater than or equal to the arrival rate of all data queues and processing queues, i.e. the mean rate stability\cite{ref21}.

\section{Problem Solution and Algorithm design}

\begin{figure}[!t]
  \centering
  \includegraphics[scale= 0.6]{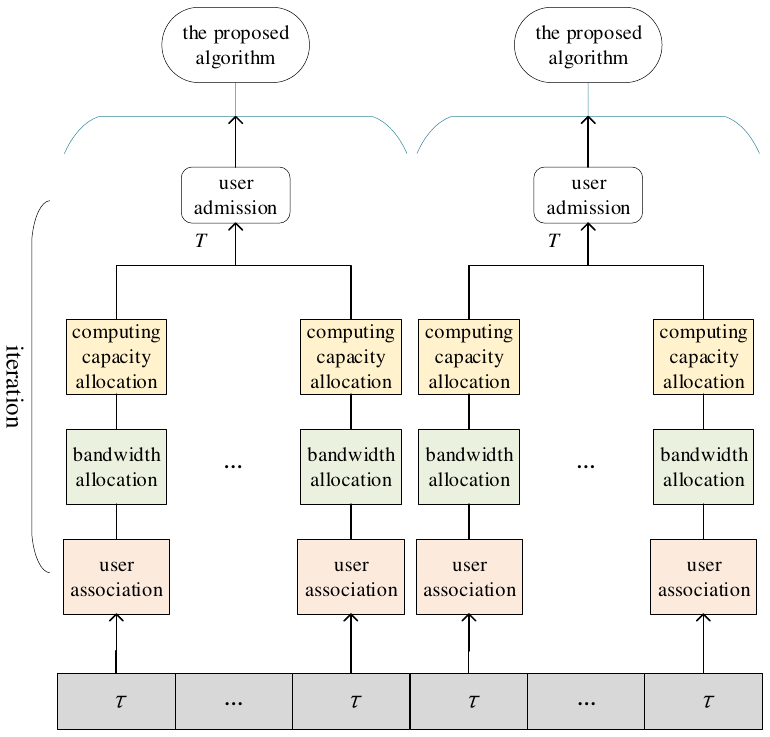}
  \hfil
  \caption{Framework of the two time-slot algorithm.}
  \label{fig_sim}
\end{figure}

Since our optimization problem \eqref{deqn20} is stochastic and complex in temporal domain and mixed in multi-time scale, we need to solve it by decomposing the two time-slot problem into many single time-slot subproblems with the help of Lyapunov framework as illustrated in Fig. 2. Besides, an algorithm for the user admission problem will be designed. Next, we will show the proposed algorithm is capable of achieving the revenue-cost trade-off in MEC systems.

\subsection{The Lyapunov Optimization-Based Algorithm}
We propose an algorithm based on the Lyapunov optimization and substitute the queue formula proposed in Section II into Lyapunov framework. Let ${\mathbf{\Theta}}(t) =[{{\mathbf{\Upphi_I}}}(t),{{\mathbf{\Upphi_{II}}}}(t),{\mathbf{\Phi}}(t)]$ be a concatenated vector, and we define the Lyapunov function as
\begin{equation}
  \label{deqn21}
  L({\mathbf{\Theta}}(t))=\frac{1}{2}\left[{{\mathbf{\Upphi_I}}}^2(t)+{{\mathbf{\Upphi_{II}}}}^2(t)+{\mathbf{\Phi}}^2(t)\right].
\end{equation}

Then the LTS conditional Lyapunov drift $\Delta_T({\mathbf{\Theta}}(l))$ is given by
\begin{equation}
  \label{deqn22}
   \Delta_T({\mathbf{\Theta}}(l))={\mathbb{E}}[L({\mathbf{\Theta}}(l+T)-{\mathbf{\Theta}}(l))|{\mathbf{\Theta}}(l)],
\end{equation}
where ${\mathbf{\Theta}}(l)=\{{{\mathbf{\Upphi_I}}}(t),{{\mathbf{\Upphi_{II}}}}(t),{\mathbf{\Phi}}(t),t\in [l,l+T-1]\}$. Then, the drift-plus-penalty expression of \eqref{deqn21} is expressed as
\begin{equation}
  \label{deqn23}
  \Delta_T({\mathbf{\Theta}}(l))-V{\mathbb{E}}\{G(l)|{\mathbf{\Theta}}(l)\}.
\end{equation}
\textbf{Remark 2.} \textit{From (\ref{deqn23}), we notice that the control parameter $V>0$ which is from normalized form of Lyapunov optimization represents the extent of drift-plus-penalty and controls the weight of penalty. In our proposed algorithm, the larger parameter $V$ increases the weight of penalty, i.e., system utility and causes a increase in the final optimization value. Hence, adjusting parameter $V$ can balance the importance of queue stability and system utility and acquire an ideal result we want.}

We derive the following theorem to provide an upper bound on the above drift-plus-penalty expression.

\textit{Theorem 1}: Suppose $G(t)$ is i.i.d. over slots. For arbitrary ${\boldsymbol{y}}(l)$, ${\boldsymbol{x}}(t)$, ${\boldsymbol{w}}(t)$, ${\boldsymbol{f}}(t)$, all parameters $V > 0$, and all possible values of ${\mathbf{\Theta}}(l)$, $\Delta_T({\mathbf{\Theta}}(l))-V{\mathbb{E}}\{G(l)|{\mathbf{\Theta}}(l)\}$ is upper bounded by \eqref{deqn24}, where $C$ meets \eqref{deqn25}.

\textit{Proof}: Please refer to Appendix A.

The stochastic optimization theory indicates that a stochastic optimization problem is solved by minimizing the upper bound of its drift-plus-penalty expression subject to the same constraints except the stability one in \cite{ref21}. Therefore, we need to minimize right-hand-side of \eqref{deqn24} to solve \eqref{deqn20} subject to (C1)-(C6), because (C7) is a stability constraint. Therefore, the original multi-time scale optimization problem for long-term revenue can be equivalently transformed into an optimization of revenue on multiple LTSs after the above process. Then, we can get the following optimization problem which is expressed by \eqref{deqn26}.

\begin{strip}
\begin{equation}
  \label{deqn24}
  \begin{split}
    &\Delta_T({\mathbf{\Theta}}(l))-V{\mathbb{E}}\{G(l)|{\mathbf{\Theta}}(l)\} \leqslant C -  {\sum_{t=pl}^{p(l+1)-1}} {{\mathbf{\Upphi_I}}}(t) {\mathbb{E}}\Bigg\{\left[\tau \boldsymbol{y}(l)\cdot \boldsymbol{r}(t)  - \boldsymbol{y}(l)\cdot \boldsymbol{A}(t)\right] \Bigg|{\mathbf{\Theta}}(l) \Bigg\} - \sum_{t=1}^{l+T-1} {{\mathbf{\Upphi_{II}}}}(t) {\mathbb{E}}\Bigg\{\Bigg[\sum_{u=1}^{U} y_u(l) B_{bus}\tau- \\
    &\boldsymbol{y}(l)\cdot \boldsymbol{r}(t)\Bigg]\Bigg|{\mathbf{\Theta}}(l) \Bigg\} - \sum_{t=1}^{l+T-1} {\mathbf{\Phi}}(t) {\mathbb{E}}\Bigg\{\Bigg[\boldsymbol{y}(l)\cdot \boldsymbol{f}(t)- \sum_{u}{F_{u}(y_u(l)B_{bus})}{(t)}\Bigg]\Bigg|{\mathbf{\Theta}}(l) \Bigg\} - V{\mathbb{E}} \Bigg\{\Bigg[G_L(l)-\eta  {\sum_{t=pl}^{p(l+1)-1}}P(t)\Bigg]\Bigg|{\mathbf{\Theta}}(l)\Bigg\}.
  \end{split}
\end{equation}

\begin{equation}
  \label{deqn25}
  \begin{split}
    C \geqslant &\frac{1}{2} \Bigg\{ {\left[ \sum_{u=1}^{U}y_u(l) {\sum_{t=pl}^{p(l+1)-1}}r_u(t)\tau \right]^2} + \Bigg[\sum_{u=1}^{U}y_u(l) {\sum_{t=pl}^{p(l+1)-1}}  A_u(t)\Bigg]^2 + {\left[ \sum_{u=1}^{U}y_u(l) {\sum_{t=pl}^{p(l+1)-1}}B_{bus}\tau \right]}^2 + \\
    & \max_{u\in U^S(l)} \Bigg[ y_u(l) {\sum_{t=pl}^{p(l+1)-1}}r_u(t) \Bigg]^2 + {\left[ \sum_{u=1}^{U}y_u(l) {\sum_{t=pl}^{p(l+1)-1}}f_u(t) \right]}^2 + \max_{u \in U^S(l)} \Bigg[  {\sum_{t=pl}^{p(l+1)-1}} {F_{u} (y_u(l)B_{bus})(t)} \Bigg]^2  \Bigg\}.
  \end{split}
\end{equation}

\begin{equation}
  \label{deqn26}
  \begin{split}
    &\max_{{\boldsymbol{y}}(l),{\boldsymbol{x}}(t),{\boldsymbol{w}}(t),{\boldsymbol{f}}(t)} [{{\mathbf{\Upphi_{II}}}}(t)- \mathbf{\Upphi_I}] \boldsymbol{y}(l)\cdot \boldsymbol{r}(t) -{{\mathbf{\Upphi_{II}}}}(t) \sum_{u}y_u{(l)}B_{bus}\tau-{\mathbf{\Phi}}(t)\boldsymbol{y}(u)\cdot \boldsymbol{f}(t)+{\mathbf{\Phi}}(t) \sum_{u} {F_{u}(y_u(l)}{B_{bus})(t)}+V\eta P(t)\\
    &\ \ \ \ \ \ \ \  s.t. \ \; (\text{C}1)-(\text{C}6).
  \end{split}
\end{equation} 
\end{strip}
From the principle of opportunistically minimizing an expectation, minimizing ${\boldsymbol{f}}(t)$ can ensure that ${\mathbb{E}}\{{\boldsymbol{f}}(t)|{\mathbf{\Theta}}(t)\}$ is minimized. Therefore for the objective function in \eqref{deqn26}, we obtain by ignoring constant $C$, ${{{\mathbf{\Upphi_I}}}}\left( t \right)\sum\limits_{u = 1}^U {{y_u}\left( l \right){A_u}\left( t \right)} $ and ${G_L}\left( l \right)$ in \eqref{deqn24} and removing the conditional expectations in \eqref{deqn24}. {Since user admission is in long time scale, we use subproblem separation to separate the long and short time scale problems for ease of processing. The subproblems in long and short time scale will be solved later by iterative integration.} Obviously, user association, bandwidth allocation and computing capacity allocation are highly coupled with each other in \eqref{deqn26}. We further decompose these optimization variables to develop low-complexity algorithms in the following subsections. 

\subsection{Solution of Resource Allocation Subproblem in Short Time Scale}
We obtain the solution of the coupled optimization problem \eqref{deqn26} by integrating the algorithms through iterative optimization. Under given user association ${\boldsymbol{x}}(t)$ and bandwidth allocation ${\boldsymbol{w}}(t)$, the computing resource allocation subproblem can be expressed by
\begin{equation}
  \label{deqn27}
  \begin{split}
    &\max_{{\boldsymbol{f}}(t)}{\mathbf{\Phi}}(t)\boldsymbol{y}(l)\cdot \boldsymbol{f}(t)-V\eta \sum_{u,k}x_{uk}(t)\kappa_{esc}f_{uk}^3(t)\\
    &s.t. \ \; (\text{C}4):f_u(t) \geqslant \frac{y_u {F_{u}(a_u(t))}}{\sum\limits_{u=1}^{U}z_{um}\tilde{t}_m - y_u\frac{a_u(t)}{B_{bus}}-y_u\frac{a_u(t)}{r_U(t)}},\forall u,t, \\
    &\qquad(\text{C}6):\sum_{u=1}^U x_{uk}(t)f_{uk}(t) \leqslant F_k, \forall k,t.
  \end{split}
\end{equation} 
where the other terms of equation \eqref{deqn26} are constants under the above condition. Obviously, the objective function of \eqref{deqn27} is {concave} and its constraints are linear, so it is a convex optimization problem. Therefore we obtain the optimized solution $f^*(t)$ directly through the convex optimization method\cite{ref22} in polynomial time using standard CVX tools\cite{ref23}.

Under given ${\boldsymbol{w}}(t)$ and ${\boldsymbol{f}}(t)$ we obtain the user association problem which is denoted as
\begin{equation}
  \label{deqn28}
  \begin{split}
    &\min_{{\boldsymbol{x}}(t)}\left[{{\mathbf{\Upphi_{II}}}}(t)-{{\mathbf{\Upphi_I}}}(t)\right]\boldsymbol{y}(u)\cdot \boldsymbol{r}(t) +V\eta \sum_{u,k}x_{uk}(t)\kappa_{esc}f_{uk}^3(t)\\
    &s.t.\ \;(\text{C}2)-(\text{C}3),
  \end{split}
\end{equation} 
which is converted to
\begin{equation}
  \label{deqn29}
  \begin{split}
    &\min_{{\boldsymbol{x}}(t)} \left[({{\mathbf{\Upphi_{II}}}}(t)-{{\mathbf{\Upphi_I}}}(t))\sum_{u,k}y_u(l)r_{uk}(t)+V\eta \sum_{u,k}\kappa_{esc}f_{uk}^3(t)\right]\\
    &\qquad \ x_{uk}(t)\\
    &s.t.\ \;(\text{C}2)-(\text{C}3),
  \end{split}
\end{equation} 
whose optimal solution is expressed as
\begin{equation}
  \label{deqn30}
  x_{uk} = 
  \begin{cases}
    1,& k = k^*,\\
    0,& k\neq k^*,
  \end{cases}
\end{equation} 
where 
\begin{equation}
  \label{deqn31}
  \begin{split}
    k^* = \text{arg}\min_{k\in K^S} \left\{ ({{\mathbf{\Upphi_{II}}}}(t)-{{\mathbf{\Upphi_I}}}(t))y_u(l) r_{uk}(t)+V \eta \kappa_{esc} f_{uk}^3(t)\right\}
  \end{split}
\end{equation} 

\begin{algorithm}[H]
  \caption{Iterative Algorithm for Resource Allocation for Given Admission.}\label{alg:alg1}
  \begin{algorithmic}[1]
   \REPEAT
   \STATE Set initial STS $t=l$ and obtain the current queue state ${{\mathbf{\Upphi_I}}}(t)$, ${{\mathbf{\Upphi_{II}}}}(t)$, and ${\mathbf{\Phi}}(t)$. 
   \STATE Set $q=0$, user association $x^0(t)$ and computing capacity allocation $f^0(t)$.
   \STATE Set $l = 0$ and the iteration constraints $ \varepsilon > 0$.
   \STATE Obtain $w^0(t)$ by solving bandwidth allocation subproblem through $x^0(t)$ and $f^0(t)$.
   \STATE Obtain the value of optimization problem \eqref{deqn26}, i.e. $N^0(t)$.
   \REPEAT
   \STATE $q = q + 1$.
   \STATE Obtain $x^q(t)$ by solving user association subproblem through $w^{q-1}(t)$ and $f^{q-1}(t)$.
   \STATE Obtain $f^q(t)$ by solving computing capacity allocation subproblem through $w^{q-1}(t)$ and $x^{q}(t)$.
   \STATE Obtain $w^q(t)$ by solving bandwidth allocation subproblem through $x^q(t)$ and $f^q(t)$.
   \STATE Obtain the value of $N^q(t)$.
   \UNTIL $\left\lvert N^q(t)-N^{q-1}(t)\right\rvert \leqslant \varepsilon$.
   \STATE $t = t + 1$.
   \STATE Update ${{\mathbf{\Upphi_I}}}(t)$, ${{\mathbf{\Upphi_{II}}}}(t)$, and ${\mathbf{\Phi}}(t)$.
   \UNTIL $t=l+T-1$.
  \end{algorithmic}
\end{algorithm}

Next, if user association ${\boldsymbol{x}}(t)$ and computing capacity allocation ${\boldsymbol{f}}(t)$ are known, then the bandwidth allocation subproblem is given by
\begin{equation}
  \label{deqn32}
  \begin{split}
    &\min_{{\boldsymbol{w}}(t)} \\
    &\sum_{u,k} [({{\mathbf{\Upphi_{II}}}}(t)-{{\mathbf{\Upphi_I}}}(t)]x_{uk}(t) \log_2\left(1+\frac{{g_{uk}(t)p_u}}{I_{uk}(t)+{\sigma^2}}\right)w_{uk}(t) \\
    &s.t.\ \;(\text{C}4)':\sum_k x_{uk}(t)w_{uk}(t)\log_2\left(1+\frac{{g_{uk}(t)p_u}}{I_{uk}(t)+{\sigma^2}}\right) \geqslant \\
    &\qquad \qquad \quad \ \frac{a_u(t)}{\tilde{t}_m -t_u^{comp}(t)-t_u^{bus}(t)},\forall u,t,\\
    &\qquad (\text{C}5):\sum_{u\in U}x_{uk}(t)w_{uk}(t)\leqslant W_k,\forall k,t.
  \end{split}
\end{equation} 
It can be seen that the objective function of the problem is a linear function, and the constraints are linear, so it is a linear programming problem, and the solution ${\boldsymbol{w}}(t)$ is directly obtained by optimization methods such as the interior point method\cite{ref24} by using standard CVX tools. 

We use the idea of the greedy algorithm to iterate the above three solutions for the subproblem and arrive at the suboptimal solution, which is summarized in Algorithm 1.

\subsection{Solution of User Admission Subproblem in Long Time Scale}

The original problem \eqref{deqn20} is decomposed into two subproblems on the time scale, one is \eqref{deqn26} in short time scale and the other is denoted as
\begin{equation}
  \label{deqn33}
  \begin{split}
    &\max_{{\boldsymbol{y}}(l)} G_L(l) - \eta G_S(l) \\
    &s.t. \ \;(\text{C}1):y_u(l)\in \{0,1\},\forall u, \\
    &\qquad (\text{C}8):\text{feasibility of problem \eqref{deqn26}}.
  \end{split}
\end{equation} 
Obviously, this problem is a nonlinear 0-1 integer programming problem, and the constraint (C8) needs to determine the solvability of \eqref{deqn26}. Therefore it can not be solved by conventional methods. According to the above condition that the random arrival process of tasks of user \textit{u} is independently and identically distributed between STSs, the constraint (C8) is transformed into (C4) and (C7). The Lyapunov architecture makes the above solution , i.e. Algorithm 2, satisfy (C7), so we solely need to transform {(C8)} into (C4) there. Then, \eqref{deqn33} is expressed as
\begin{equation}
  \label{deqn34}
  \begin{split}
    &\max_{{\boldsymbol{y}}(l)} \boldsymbol{v}(l)\cdot \boldsymbol{y}(l)  - \eta  {\sum_{t=pl}^{p(l+1)-1}}P(t) \\
    &s.t. \ \;(\text{C}1):y_u(l)\in \{0,1\},\forall u, \\
    &\qquad (\text{C}4):y_u(l)t_u(t) \leqslant \sum_{m=1}^M z_{um}(t)\tilde{t}_m,\forall u,t\in [l,l+T-1].
  \end{split}
\end{equation} 

The optimization problem \eqref{deqn34} is a linear 0-1 integer programming problem and is solved by iterating with \eqref{deqn26}. The whole procedure is shown in Algorithm 2.

\begin{algorithm}[H]
  \caption{User Admission Algorithm.}\label{alg:alg2}
  \begin{algorithmic}[1]
    \STATE Set initial LTS \textit{l}.
    \REPEAT
    \STATE Set $j=0$, and determine the initial $y_u^0(l)$ that satisfies the condition.
    \STATE Solve \eqref{deqn26} through Algorithm 1 and obtain ${\boldsymbol{x}}(t)$, ${\boldsymbol{w}}(t)$ and ${\boldsymbol{f}}(t)$.
    \STATE $j = j + 1$.
    \STATE Obtain the iteration consequence $y_u^j(l)$ by solving \eqref{deqn34}.
    \STATE Calculate the value of $G(l)$.
    \UNTIL $G(l)$ no longer changes significantly.
    \STATE $l = l + 1$.
  \end{algorithmic}
\end{algorithm}

\subsection{Dynamic Solution to the Optimization Problem}
As mentioned above, we decompose the original complex stochastic optimization problem \eqref{deqn20} into two subproblems. As shown in Fig.2, we propose a short-time-slot resource allocation algorithm and a long-time-slot user admission algorithm to solve it. Due to the specificity of the two time-slot iteration algorithm, we need to solve the problem offline and adopt the strategies online. The detailed procedure to solve \eqref{deqn20} is summarized in Algorithm 3.

\begin{algorithm}[H]
  \caption{Multi-time Scale User Admission and Resource Allocation algorithm.}\label{alg:alg3}
  \begin{algorithmic}[1]
  \STATE In each LTS $l = iT,i=0,1,....$ obtain the current queue state ${{\mathbf{\Upphi_I}}}(t_k)$.
  \STATE Determine the user admission $y_u(l)$ by calling Algorithm 2.
  \STATE In each STS $t\in [l,l+T]$, obtain user association ${\boldsymbol{x}}(t)$, bandwidth allocation ${\boldsymbol{w}}(t)$ and computing capacity allocation ${\boldsymbol{f}}(t)$ according to \eqref{deqn26} by calling Algorithm 1.
  \STATE $t = t + 1$.
  \STATE Update the current queue value ${{\mathbf{\Upphi_I}}}(t)$, ${{\mathbf{\Upphi_{II}}}}(t)$, and ${\mathbf{\Phi}}(t)$ according to \eqref{deqn7}, \eqref{deqn14} and \eqref{deqn15}.
  \end{algorithmic}
  \label{alg3}
\end{algorithm}

\subsection{Analysis of the Proposed Algorithms}
In this subsection, we analyze the temporal computational complexity and the convergence of the proposed algorithms.

In Algorithm 1, we adopt alternating iteration of three subproblems and obtain the solutions in closed form by convex optimization. According to the greedy algorithm and convex optimization theory\cite{ref25}, iteration of three convex subproblems can ensure $|N^q(t)-N^{q-1}(t)|\leqslant \varepsilon$ quickly. Therefore, the convergence of Algorithm 1 is obvious but only sub-optimality is guaranteed\cite{ref5}. From the perspective of complexity, the complexity of \eqref{deqn27}, \eqref{deqn28} and \eqref{deqn32} are ${\mathcal{O}}(U^3K)$, ${\mathcal{O}}(UK)$ and ${\mathcal{O}}((UK)^{3.5}))$\cite{ref25}. We assume the number of iterations is $L_1$, then the complexity of Algorithm 1 is ${\mathcal{O}}((U^3K+UK+(UK)^{3.5})L_1)$. This decoupled algorithm can make good use of convex optimization methods to solve complex coupling problem \eqref{deqn26}.

In Algorithm 2, since \eqref{deqn34} a linear 0-1 integer programming problem and iterating with \eqref{deqn26}, we assume the number of iterations is $L_2$, then the complexity is ${\mathcal{O}}((U^2+\gamma_1)L_2)$ where $\gamma_1$ is the complexity of Algorithm 1. Obviously, the convergence of the Algorithm 2 depends on Algorithm 1 since the iteration is performed with it. Therefore, the proposed algorithm can reach convergence after several iterations as Algorithm 1 converges quickly. Algorithm 3 indicates that above algorithms run on time slots, therefore at LTS \textit{l} the complexity of Algorithm 3 is ${\mathcal{O}}((U^2+\gamma_1)L_2p)$. In the way, the complex stochastic optimization problem \eqref{deqn20} is decomposed into low-complexity subproblems iteratively solved. 

\section{Simulation Result}
In this section, we first set the simulation parameters and then demonstrate simulation results to evaluate the performance of the proposed algorithms. 

\subsection{Simulation Parameters}
As mentioned above, we consider system level simulation of uplink in a small cell F-RAN according to a 3GPP normative document of small cell network\cite{ref26}. Four SBSs are deployed in a small cell area, with a total coverage area of $200\text{m} \times 200\text{m}$. The SBSs provide admission and resource allocation for users. We model the path loss of radio access link of the small cell network and use a hexagonal cellular deployment model. The distance between the user and the SBS is within the standard of the 3GPP document and only the outdoor access link exists. At STS \textit{t}, let $d_{uk}(t)$ be the distance between SBS \textit{k} and user \textit{u}, and users all moves randomly within the area at a speed of 3km/h. Note that the path loss between SBS \textit{k} and user \textit{u} is dependent on the link state of LoS and NLoS. When it is a LoS link, the path loss is given by
\begin{equation}
  \begin{split}
    \label{deqn1}
    \mu^{LoS/NLoS}_{uk}(t) = &22.0{\log}_{10}(d_{uk}(t))+28.0+\\
    &20{\log}_{10}(F^q),
  \end{split}
\end{equation}
and when it is a NLoS link, the path loss is given by
\begin{equation}
  \begin{split}
    \label{deqn2}
    \mu^{LoS/NLoS}_{uk}(t) = &36.7{\log}_{10}(d_{uk}(t))+22.7+\\
    &26{\log}_{10}(F^q),
  \end{split}
\end{equation}
where $F^q$ indicates the carrier frequency. The LoS probability that determines the LoS/NLoS link state is denoted as
\begin{equation}
  \label{deqn3}
  p^{LoS}_{uk}(t) = \min\left(\frac{18}{d_{uk}(t)},1\right)\left(1-e^{-\frac{d_{uk}(t)}{36}}\right)+e^{-\frac{d_{uk}(t)}{36}},
\end{equation}
therefore the NLoS probability is $p^{NLoS}_{uk}(t) = 1-p^{LoS}_{uk}(t)$. Then the channel gain is expressed as
\begin{equation}
  \label{deqn4}
  g_{uk}(t) = \left(p^{LoS}_{uk}10^{\mu^{LoS}_{uk}}+p^{NLoS}_{uk}10^{\mu^{NLoS}_{uk}}\right)^{-1}.
\end{equation}

In our proposed algorithm, we notice that parameter $p$ of $T=p\tau$ in our algorithm represents the relationship between the length of LTSs and STSs. Hence, too small $p$ will affect the effect of multi-time scales and optimization of the algorithm and too large $p$ will increase the running time of our algorithm but the improved algorithm performance is not significant. Therefore, we set appropriate values of $T$ and $\tau$. Part of simulation parameters are summarized in Table II\cite{ref27}.

\begin{table}[h]
  \renewcommand\arraystretch{1.5}
  \caption{Part of Simulation Parameters}
  \centering
  \begin{tabular}{p{4cm}<{\centering}|p{2cm}<{\centering}}
  \hline
  \hline
  Parameter & Value \\
  \hline
  Long Time Slot (LTS), $T$ & 1 s\\
  \hline
  Short Time Slot (STS), $\tau$ & 0.1 s\\
  \hline
  Transmit power, $p_u$ & 37 dBm\\
  \hline
  Bandwidth resource, $W_k$ & 10 MHz\\
  \hline
  Computing capacity, $F_k$ & 200 Gigacycle/s\\
  \hline
  The noise power, {$\sigma^2$} & -100 dBm\\
  \hline
  The carrier frequency, $F^q$ & 3.5 GHz\\
  \hline
  Bus bandwidth, $B_{bus}$ & 10 Gbps\\
  \hline
  Arrival rate, $\lambda$ & 50\\
  \hline
  The total number of users, $U$ & 60\\
  \hline
  Adjusting parameter, $\eta$ & $10^{-6}$\\
  \hline
  The number of iterations, $L_1$ & 50\\
  \hline 
  \hline
  \end{tabular}
\end{table}

According to the computing delay requirements of some services of ultra reliable low latency communications and combining task scenarios\cite{ref28,ref29}, we set the basis delay requirement as 20 ms. Furthermore, we assume three task types whose delay requirements and model parameters progressive by task type and are equally distributed to the task set. The delay limits and model parameters of the tasks are shown in Table III.
\begin{table}[h]
  \renewcommand\arraystretch{1.5}
  \caption{Task Parameters}
  \centering
  \begin{tabular}{p{1.2cm}<{\centering}|p{1.5cm}<{\centering}|p{1.2cm}<{\centering}}
    \hline
    \hline
    Type & $\tilde{t}_m$ & $n_m$\\
    \hline
    1 & 20 ms & 1 \\
    \hline
    2 & 40 ms & 3 \\
    \hline
    3 & 60 ms & 5\\
    \hline
    \hline
  \end{tabular}
\end{table}

\subsection{Convergence of the Proposed Algorithm}
\begin{figure}[!t]
  \centering
  \includegraphics[scale=0.57]{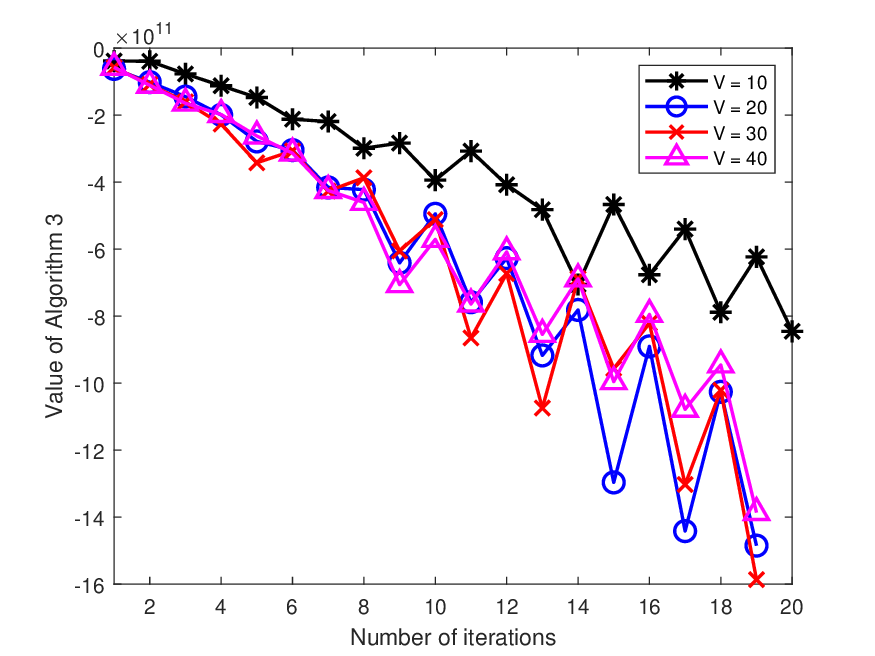}
  \caption{Convergence of the proposed Algorithm}
  \label{fig_2}
\end{figure}

For convenience, we integrate the value of STS iteration and the value LTS iteration, i.e., the system revenue and cost at each STS, to show our overall convergence. Fig. \ref{fig_2} shows the convergence of the proposed algorithm under different parameters $V$. From Fig. \ref{fig_2}, we can see that the convergence of the proposed Algorithm is fast and the trend is basically fixed after convergence. Since Algorithm 1 is nested into Algorithm 2 for computing, Algorithm 1 will stop as system utility of Algorithm 2 converges and the trend will be basically fixed after convergence. Also, a larger $V$ indicates a larger penalty weight in Lyapunov drift plus penalty which increases the weight of $G_S(l)$ relative to the overall queue stability and thus increases the impact of power consumption, therefore it leads to a change in the value of the utility which increases and a more volatile fluctuation after convergence as $V$ becomes larger, which is a good proof of \textbf{Remark 2}.

\subsection{Performance of the Proposed Algorithm}
To verify the performance of the proposed algorithm, we will consider the following schemes:
\begin{itemize}
  \item{Fixed Allocation (FA): The scheme only optimizes user admission.}
  \item{Fixed Channel (FC): The scheme is that user association and bandwidth allocation are fixed and optimize access variables and computational resource allocation.}  
  \item{Traditional Computing (TC)\cite{ref5}: The scheme allocates computing resources based on input data size according to the traditional computing model.} 
\end{itemize}

\begin{figure}[!t]
  \centering
  \includegraphics[scale=0.57]{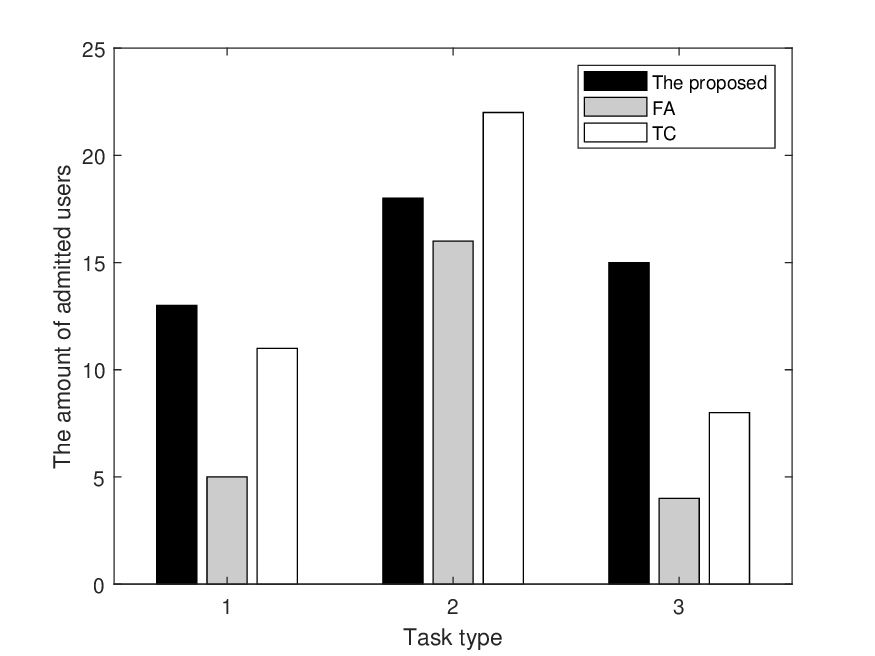}
  \caption{The amount of admitted users varying with task types.}
  \label{fig_3}
\end{figure}
We demonstrate the amount of admitted users varying with three different task types in Fig. \ref{fig_3}. As we can see, different values of $\widetilde{t}_m$ affect the amount of admitted users. From \eqref{deqn16}, we notice that $\widetilde{t}_m$ affects the system revenue and can cause that our algorithm will choose more valuable users. This characteristic is also shown in figure and the amount of admitted users of task 2 is more than other tasks. In our system, integrated user value is affected by {the required computing resource} $F_{um}$ besides $\widetilde{t}_m$ and therefore users of task 2 become the most valuable in our parameters. However, our proposed algorithm can handle the difference between three types of tasks well and make user admission stable as we can see in Fig. \ref{fig_3} compared with other contrast algorithms. 

\begin{figure}[!t]
  \centering
  \includegraphics[scale=0.57]{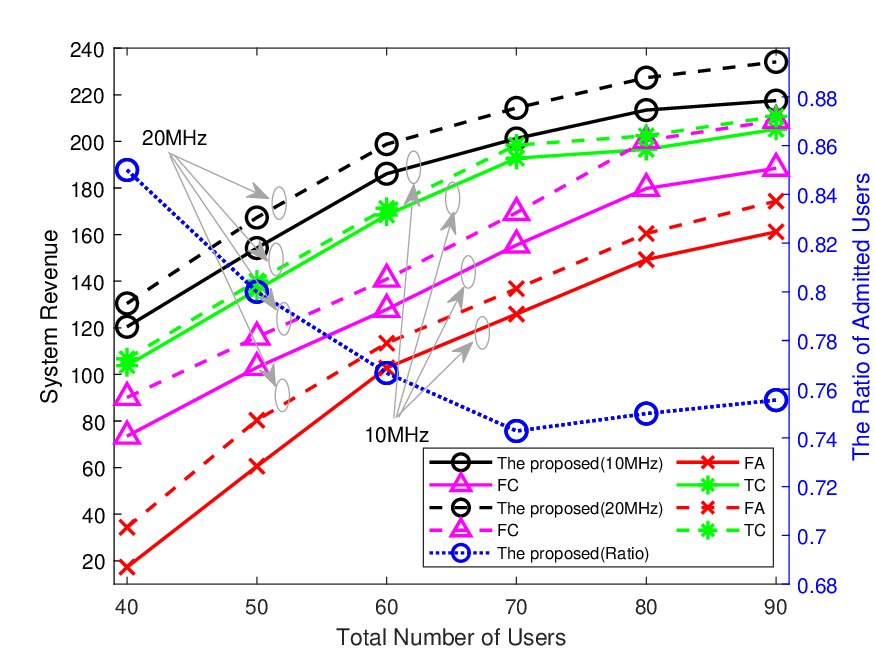}
  \caption{{System utility varying with total number of users.}}
  \label{fig_4}
\end{figure}

\begin{figure}[!t]
  \centering
  \includegraphics[scale=0.57]{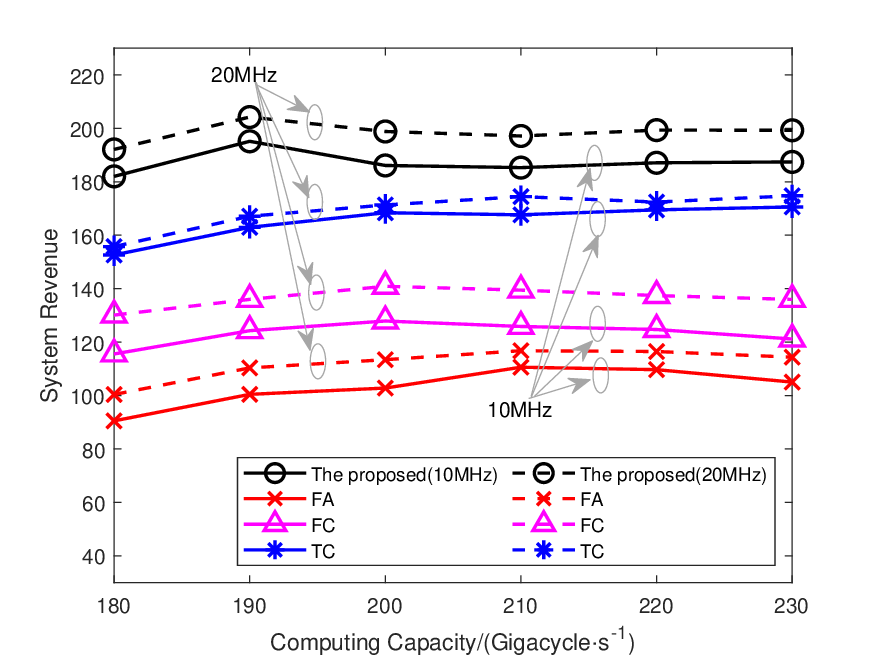}
  \caption{{System utility varying with computing capacity.}}
  \label{fig_5}
\end{figure}

\begin{figure}[!t]
  \centering
  \includegraphics[scale=0.57]{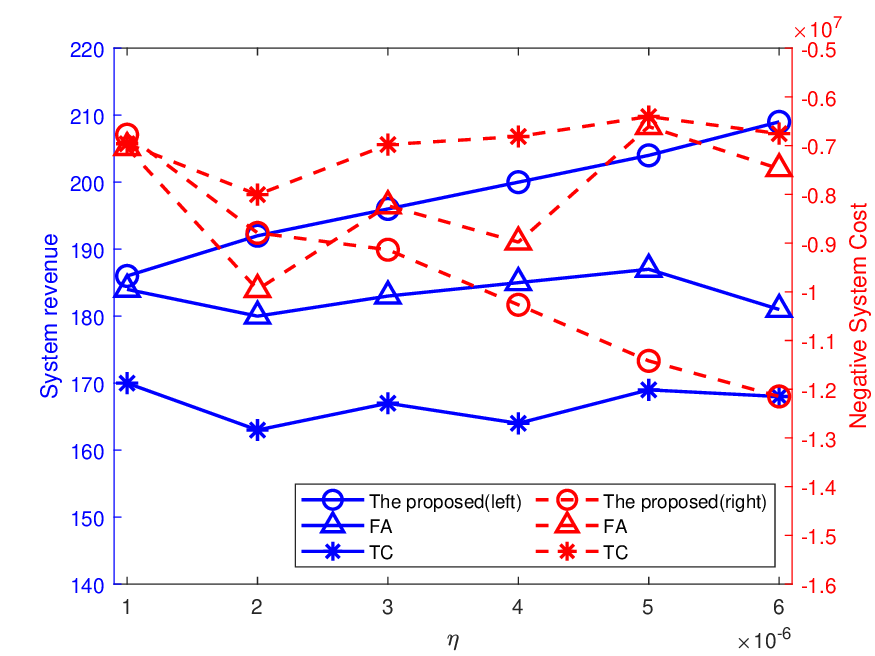}
  \caption{Trade-off between system revenue and system cost vs. $\eta$.}
  \label{fig_6}
\end{figure}
The characteristics of system utility with total number of users under different bandwidth values as Fig. 5 shows, it is found that it increases with the total number of users and slower after the total number of users is greater than 70. To better indicate the momentum of the system utility, we add the blue dotted line which represent the ratio of admitted users of our proposed algorithm as right vertical axis in Fig. 5 shows. When the total number of users is small, the resources are sufficient and resource allocation can be efficient, so the system utility will increase quickly as the total number of users increases. However, when the total number of users is large, the resources are limited and resouce alloction will not be efficient, therefore the rising tendency of utility will be reduced. The radio of admitted users will decrease first and then become stable after the total number of users is 70, which shows the increasing momentum of system utility will become slow when the total number of users become large. The comparison algorithms all have this property, but the trend is different for different algorithms. To make figure concise as far as possible, we do not show the ratio of admitted users of the comparison algorithms which has been compared in Fig. 4. There is a slight but insignificant increase in system utility for higher bandwidth values, therefore the bandwidth value change has a small impact on the system.

We compare system utility with computing capacity under different bandwidth values in Fig. 6, and it is seen that there is a maximum value in $F_k = 190$ Gigacycle/s. This is because system utility depends on the number of association users and power consumption and our algorithm needs to make a trade-off between them. When $F_k$ reaches a certain amount, our proposed algorithm can take it to a better trade-off utility value. However when $F_k$ continues to rise, the fairness design of this algorithm will allow more users to admit and thus generate more power consumption. Therefore it will lead to a decrease in the value of utility. While $F_k$ increases to a certain magnitude, the number of associated users will not continue to increase due to bandwidth resource, so the change is no longer significant. This property is also present in comparison algorithms, mainly because the utility defined in this paper are indirectly influenced by the allocation of multiple resources. The increase in computing resource has little effect on total utility, so its change is insignificant and even has the characteristic of decreasing with its increase. We can see in this graph again that higher bandwidth values do not have a significant impact on total utility.

We plot the trade-off between system revenue and system cost vs. $\eta$ in Fig. \ref{fig_6}. The system utility \eqref{deqn18} indicate that our algorithm can balance system revenue and cost, and in this figure we can see that our proposed algorithm can get better trade-off than comparison algorithm while $\eta$ increases, which verifies \textbf{Remark 1}. There is obvious increase on system revenue and decrease on negative system cost when $\eta$ increases in our proposed algorithm but this trend is not that significant in other comparison, which means our proposed algorithm significantly attach importance to user admission and stabilize the system utility. Therefore, the system utility is increased by balancing revenue at LTSs and cost at STSs in the proposed algorithm and make systems increasingly stable compared with other algorithms.  

\section{Conclusion}
In this paper, we studied the dynamical resource allocation problem of specific characteristic of tasks in MEC systems. Specially, the stochastic optimization problem we proposed was decomposed into user admission at LTS and resource allocation at STS by the Lyapunov optimization technique and we decoupled the optimization variables for efficient algorithm design and solve each subproblem at low complexity. Simulation results has demonstrated that, compared with the benchmarks, the proposed algorithm improves the performance of user admission and resource allocation efficiently and achieves a flexible trade-off between system revenue and cost at multi-time scales and considering semantic extraction tasks.

{\begin{strip}
\appendix[{{A Proof of Theorem 1}}]

First of all, we have that ${\left\{ {\max \left[ {A - B,0} \right] + C} \right\}^2} \le {A^2} + {B^2} + {C^2} - 2A\left( {B - C} \right)$ always holds {if $A\geq 0$, $B\geq 0$ and $C \geq 0$}. Suppose $V>0$, then squaring both sides of \eqref{deqn7} yields
\begin{align}
    &{{\mathbf{\Upphi_I}}}(l+T-1)^2 \leqslant {{\mathbf{\Upphi_I}}}(t)^2+\left[\sum_{t=pl}^{p(l+1)-1} \tau \boldsymbol{y}(l) \cdot \boldsymbol{r}(t) \right]^2 + \left[ \sum_{u=1}^U y_u(l) {\sum_{t=pl}^{p(l+1)-1}}A_u(t)\right]^2 - 2 {\sum_{t=pl}^{p(l+1)-1}}{{\mathbf{\Upphi_I}}}(t)\boldsymbol{y}(l)\cdot\left[\tau  \boldsymbol{r}(t) - \boldsymbol{A}(t) \right], 
\end{align}
For \eqref{deqn14}, similarly, we have
\begin{align}
    {{\mathbf{\Upphi_{II}}}}(l+T-1)^2 \leqslant &{{\mathbf{\Upphi_{II}}}}(t)^2+\left[\sum_{u,k}y_u(l) {\sum_{t=pl}^{p(l+1)-1}}B_{bus}\tau \right]^2 + \max_{u\in U^S(l)} \left[y_u(l) {\sum_{t=pl}^{p(l+1)-1}}r_{u}(t) \right]^2 - \notag \\ 
    & 2 {\sum_{t=pl}^{p(l+1)-1}}{{\mathbf{\Upphi_{II}}}}(t) \left[\sum_{u,k}y_u(l)B_{bus}\tau - \boldsymbol{y}(l) \cdot \boldsymbol{r}(t) \right].
\end{align}

For \eqref{deqn15}, we also have
\begin{equation}
  \label{deqn_ap3}
  \begin{aligned}
    {\mathbf{\Phi}}(l+T-1)^2 \leqslant & {\mathbf{\Phi}}(t)^2+\left[\sum_{t=pl}^{p(l+1)-1} \boldsymbol{y}(l)\cdot \boldsymbol{f}(t) \right]^2 + \max_{u\in U^S(l)} \left[ {\sum_{t=pl}^{p(l+1)-1}}\sum_{m=1}^M z_{um}(t)F_{um}(y_u(l)B_{bus}) \right]^2 - \\
    & 2 {\sum_{t=pl}^{p(l+1)-1}} {\mathbf{\Phi}}(t)\left[ \boldsymbol{y}(l)\cdot \boldsymbol{f}(t) - \sum_{u} {F_{u}(y_u(l)B_{bus})(t)} \right].
  \end{aligned}
\end{equation}

By organizing the above inequalities, we obtain
\begin{align}
    &\frac{{{\mathbf{\Upphi_I}}}(l+T-1)^2-{{\mathbf{\Upphi_I}}}(t)^2}{2} \leqslant \frac{1}{2}\Bigg\{ \Bigg[\sum_{t=pl}^{p(l+1)-1} \tau \boldsymbol{y}(l) \cdot \boldsymbol{r}(t) \Bigg]^2+ \left[ \sum_{u=1}^U y_u(l) {\sum_{t=pl}^{p(l+1)-1}}A_u(t)\right]^2\Bigg\} -  {\sum_{t=pl}^{p(l+1)-1}}{{\mathbf{\Upphi_I}}}(t) \boldsymbol{y}(l)\cdot\left[\tau  \boldsymbol{r}(t) - \boldsymbol{A}(t)\right] ,
\end{align}

\begin{align}
    \frac{{{\mathbf{\Upphi_{II}}}}(l+T-1)^2-{{\mathbf{\Upphi_{II}}}}(t)^2}{2}\leqslant &\frac{1}{2}\Bigg\{\Bigg[\sum_{u,k}y_u(l) {\sum_{t=pl}^{p(l+1)-1}}B_{bus} \tau \Bigg]^2 + \max_{u\in U^S(l)} \left[y_u(l) {\sum_{t=pl}^{p(l+1)-1}}r_{u}(t) \right]^2\Bigg\} - \notag \\ 
    &{\sum_{t=pl}^{p(l+1)-1}} {{\mathbf{\Upphi_{II}}}}(t)\left[\sum_{u,k} y_u(l)B_{bus}\tau - \boldsymbol{y}(l) \cdot \boldsymbol{r}(t) \right].
\end{align}

and 
\begin{align}
    \frac{{\mathbf{\Phi}}(l+T-1)^2-{\mathbf{\Phi}}(t)^2}{2} \leqslant &\frac{1}{2}\Bigg\{ \Bigg[\sum_{t=pl}^{p(l+1)-1}  \boldsymbol{y}(l)\cdot \boldsymbol{f}(t)\Bigg]^2 + \max_{u\in U^S(l)} \left[ {\sum_{t=pl}^{p(l+1)-1}} {F_{u}(y_u(l)B_{bus})(t)} \right]^2\Bigg\} - \notag \\ 
    &{\sum_{t=pl}^{p(l+1)-1}}{\mathbf{\Phi}}(t) \left[\boldsymbol{y}(l)\cdot \boldsymbol{f}(t)- \sum_{u}{F_{u}(y_u(l)B_{bus})(t)} \right].
\end{align}

Summing the above three equations yields
\begin{align}
    L({\mathbf{\Theta}}(l+T)-{\mathbf{\Theta}}(l)) \leqslant &\frac{1}{2} \Bigg\{ {\left[  \sum_{t=pl}^{p(l+1)-1}\tau \boldsymbol{y}(l)\cdot \boldsymbol{r}(t) \right]^2} + \Bigg[\sum_{u=1}^{U}y_u(l) {\sum_{t=pl}^{p(l+1)-1}} A_u(t)\Bigg]^2 + {\left[ \sum_{u=1}^{U}y_u(l) {\sum_{t=pl}^{p(l+1)-1}}B_{bus}\tau \right]}^2 + \notag \\
    &\max_{u\in U^S(l)} \Bigg[ y_u(l) {\sum_{t=pl}^{p(l+1)-1}}r_u(t) \Bigg]^2 + {\left[ \sum_{t=pl}^{p(l+1)-1} \boldsymbol{y}(l) \cdot \boldsymbol{f}(t) \right]}^2 + \max_{u \in U^S(l)} \Bigg[ {\sum_{t=pl}^{p(l+1)-1}} {F_{u}(y_u(l)B_{bus})(t)}\Bigg]^2  \Bigg\} -  \notag \\
    & {\sum_{t=pl}^{p(l+1)-1}}{{\mathbf{\Upphi_I}}}(t)\left[ \tau \boldsymbol{y}(l)\cdot \boldsymbol{r}(t) - \boldsymbol{y}(l) \cdot \boldsymbol{A}(t) \right] -  {\sum_{t=pl}^{p(l+1)-1}}{{\mathbf{\Upphi_{II}}}}(t)\left[\sum_{u,k}y_u(l)B_{bus}\tau- \boldsymbol{y}(l)\cdot \boldsymbol{r}(t) \right] - \notag \\
    & {\sum_{t=pl}^{p(l+1)-1}} {\mathbf{\Phi}}(t)  \left[ \boldsymbol{y}(l)\cdot \boldsymbol{f}(t) - \sum_{u} {F_{u}(y_u(l)B_{bus})(t)} \right].
\end{align}

We take conditional expectation to the above inequality and can obtain

\begin{equation}
  \label{deqn_ap4}
  \begin{split}
    &\Delta_T({\mathbf{\Theta}}(l))-V{\mathbb{E}}\{G(l)|{\mathbf{\Theta}}(l)\} \leqslant C -  {\sum_{t=pl}^{p(l+1)-1}} {{\mathbf{\Upphi_I}}}(t) {\mathbb{E}}\Bigg\{\left[\tau \boldsymbol{y}(l)\cdot \boldsymbol{r}(t)  - \boldsymbol{y}(l)\cdot \boldsymbol{A}(t)\right] \Bigg|{\mathbf{\Theta}}(l) \Bigg\} - \sum_{t=1}^{l+T-1} {{\mathbf{\Upphi_{II}}}}(t) {\mathbb{E}}\Bigg\{\Bigg[\sum_{u=1}^{U} y_u(l) B_{bus}\tau- \\
    &\boldsymbol{y}(l)\cdot \boldsymbol{r}(t)\Bigg]\Bigg|{\mathbf{\Theta}}(l) \Bigg\} - \sum_{t=1}^{l+T-1} {\mathbf{\Phi}}(t) {\mathbb{E}}\Bigg\{\Bigg[\boldsymbol{y}(l)\cdot \boldsymbol{f}(t)- \sum_{u}{F_{u}(y_u(l)B_{bus})}{(t)}\Bigg]\Bigg|{\mathbf{\Theta}}(l) \Bigg\} - V{\mathbb{E}} \Bigg\{\Bigg[G_L(l)-\eta  {\sum_{t=pl}^{p(l+1)-1}}P(t)\Bigg]\Bigg|{\mathbf{\Theta}}(l)\Bigg\},
  \end{split}
\end{equation}

where
\begin{equation}
  \label{deqn_ap5}
  \begin{split}
    &C \geqslant \frac{1}{2} \Bigg\{ {\left[ \sum_{u=1}^{U}y_u(l) {\sum_{t=pl}^{p(l+1)-1}}r_u(t)\tau \right]^2} + \Bigg[\sum_{u=1}^{U}y_u(l) {\sum_{t=pl}^{p(l+1)-1}}  A_u(t)\Bigg]^2 + {\left[ \sum_{u=1}^{U}y_u(l) {\sum_{t=pl}^{p(l+1)-1}}B_{bus}\tau \right]}^2 + \max_{u\in U^S(l)} \Bigg[ y_u(l) {\sum_{t=pl}^{p(l+1)-1}}  \\
    &\ \ \ \ \ \  r_u(t) \Bigg]^2 + {\left[ \sum_{u=1}^{U}y_u(l) {\sum_{t=pl}^{p(l+1)-1}}f_u(t) \right]}^2 + \max_{u \in U^S(l)} \Bigg[  {\sum_{t=pl}^{p(l+1)-1}} {F_{u} (y_u(l)B_{bus})(t)} \Bigg]^2  \Bigg\}.
  \end{split}
\end{equation}
Then we complete the proof of Theorem 1.
\end{strip}
}


\vfill

\end{document}